\documentclass[aps,prb,twocolumn,superscriptaddress,showpacs,floatfix]{revtex4}
\usepackage{graphicx}
\usepackage{epsfig}
\usepackage{amsmath}
\usepackage{amssymb}

\begin{document}


\title{Spin-Dynamical Analysis of Supercell Spin Configurations}

\author{Andrew L. Goodwin}
\affiliation{Department of Earth Sciences, Cambridge University,
Downing Street, Cambridge CB2 3EQ, U.K.}

\author{Martin T. Dove}
\affiliation{Department of Earth Sciences, Cambridge University,
Downing Street, Cambridge CB2 3EQ, U.K.}

\author{Matthew G. Tucker}
\affiliation{ISIS Facility, Rutherford Appleton Laboratory, Chilton,
Didcot, Oxfordshire OX11 0QX, U.K.}

\author{David A. Keen}
\affiliation{ISIS Facility, Rutherford Appleton Laboratory, Chilton,
Didcot, Oxfordshire OX11 0QX, U.K.} \affiliation{Department of Physics,
Oxford University, Clarendon Laboratory, Parks Road, Oxford OX1 3PU,
U.K.}

\date{\today}
\begin{abstract}

A model-independent approach capable of extracting spin-wave
frequencies and displacement vectors from ensembles of supercell spin
configurations is presented. The method is appropriate for those
systems whose spin-dynamical motion is well characterised by
small-amplitude fluctuations that give harmonic spin waves. The
generalised spin coordinate matrix---a quantity that may be calculated
from the observed spin orientations in an ensemble of spin
configurations---is introduced and its relationship to the
spin-dynamical matrix established. Its eigenvalues are subsequently
shown to be related to the spin-wave mode frequencies, allowing the
extraction of spin-wave dispersion curves from configurational
ensembles. Finally, a quantum-mechanical derivation of the same results
is given, and the method applied as a case study to spin Monte-Carlo
configurations of a 3D Heisenberg ferromagnet.

\end{abstract}


\maketitle

\section{Introduction}

``Atomistic'' simulations of spin interactions in magnetic systems
provide a useful mechanism of understanding both structural and
dynamical properties of this important class of materials. The
reverse Monte-Carlo (RMC) refinement method is a relevant example: a
data driven approach, RMC attempts to account for observed
experimental data in terms of ergodic assemblies of atomistic and
spin configurations.~\cite{Keen_1991,Mellergard_2000,Goodwin_2006}
There are a number of established methods of analysing such
configurations so as to calculate important thermodynamic
quantities;~\cite{Landau_1999} however, to the best of our knowledge
a method of calculating spin-wave dispersion curves in this way has
not yet been reported. In this paper, we describe one such method.
An important feature of this approach is that spin-dynamical
quantities might be extracted from a variety of experimental
methods, if these are used to generate the configurations (e.g.\ by
using RMC). Indeed, we have recently applied this approach to
reverse Monte-Carlo configurations, allowing extraction of spin-wave
dispersion curves from neutron diffraction
data.~\cite{Goodwin_preprint} Our method is appropriate for those
systems whose spin-dynamical motion is characterised by
small-amplitude fluctuations that give harmonic spin waves. We
require harmonicity only at constant temperature; some degree of
anharmonic behaviour is allowed in the sense that it is possible to
observe changes in spin-wave energies across analyses corresponding
to different sample temperatures.

Our paper is arranged as follows. In section II, we first briefly
review the well-established concept of the spin dynamical matrix,
re-casting some of the key equations in an appropriate form for our
supercell configurational analysis. We then introduce the spin
coordinate matrix, which is assembled from the normal mode spin
coordinates. We proceed by using this matrix to establish a
relationship between the observed orientations in real-space spin
configurations and the spin-wave frequencies. Furthermore, we show
explicitly how one can then extract spin-wave dispersion curves and
spin-wave mode assignments from ensembles of spin configurations,
\emph{without any prior knowledge of the exchange constants}.
Finally, we present a derivation of the key results from a
quantum-mechanical perspective. In section~\ref{casestudy}, we
illustrate the capabilities of the approach through the analysis of
configurations generated using a simple spin Monte-Carlo simulation.
We show that the method yields the expected spin-wave dispersion
relation. Finally, section IV discusses the implementation and some
possible extensions of the method.

\section{Theory}

\subsection{Semi-classical derivation of the ``spin coordinate matrix''}

Our theoretical approach builds on the well-established semi-classical
theory of spin dynamics (as described, for example in
Refs.\,\onlinecite{Keffer_1966,Kittel_1987,Ashcroft_1976}). Throughout
our analysis we will consider only systems with a single spin-alignment
axis; however, an extension to multiple-axis systems would be
straightforward. Our starting point is the spin-dynamical matrix
$\boldsymbol\Delta(\mathbf k)$, which stores dynamical information in
terms of the Heisenberg exchange integrals $J$. Its rows and columns
are indexed by the spin types $j$, and the corresponding elements are
given explicitly by the expression:~\cite{Keffer_1966}
\begin{equation}
\begin{split}
\Delta_{j,j^{\prime}}(\mathbf
k)=&-2\sum_\ell\left[\vphantom{\sum_{j^{\prime\prime}}}\frac{\sqrt{S_jS_{j^{\prime}}}}{\hbar}J{j\,
j^{\prime}\choose
0\,\ell\hphantom{\prime}}\right.\\&\quad\times\exp\{{\rm i}\mathbf
k\cdot[\mathbf r(j^{\prime}\ell)-\mathbf
r(j0)]\}\\&\left.\qquad-\delta(j,j^{\prime})\sum_{j^{\prime\prime}}\frac{S_{j^{\prime\prime}}}{\hbar}J{j\,
j^{\prime\prime}\choose
0\,\ell\hphantom{\prime\prime}}\right].\label{deltak}
\end{split}
\end{equation}
Here, the $S_j$ and $\mathbf r(j\ell)$ are the spin quantum numbers
and average positions, respectively, of each spin $j$ in each unit
cell $\ell$. The eigenvalues of the spin-dynamical matrix are the
spin-wave frequencies $\omega(\mathbf k,\nu)$, and the right
eigenvectors describe the normal mode spin-wave
displacements.~\cite{Keffer_1966}

We now construct the matrix $\boldsymbol\Sigma(\mathbf k)$, which we
will term the ``spin coordinate matrix''. It is built from the
individual spin orientations produced by the set of spin-wave modes
acting on a spin configuration. Two important results will be derived:
first, that the spin coordinate matrix is related to the spin-dynamical
matrix by the appealing relationship $\boldsymbol\Delta(\mathbf
k)\cdot\boldsymbol\Sigma(\mathbf k)=k_{\rm B}T$ (in the classical
limit); second, that the eigenvalues of $\boldsymbol\Sigma(\mathbf k)$
are the quantities $k_{\rm B}T/\omega(\mathbf k,\nu)$. The latter
property is the key result of this paper, as the spin coordinate matrix
can be calculated directly from spin configurations, and consequently
the spin-wave frequencies can be determined via matrix diagonalisation.

We begin by introducing the magnon variables $\tau(j,\mathbf k,t)$,
$\tau^\ast(j,\mathbf k,t)$. These are defined in terms of the
standard spin oscillators $\sigma^+$ and $\sigma^-$:
\begin{eqnarray*}
\tau(j,\mathbf k,t)&=&\sqrt{\frac{\hbar
S_j}{2N}}\sum_\ell\sigma^+(j\ell,t)\exp[{\rm i}\mathbf k\cdot
\mathbf r(j\ell)],\\
\tau^\ast(j,\mathbf k,t)&=&\sqrt{\frac{\hbar
S_j}{2N}}\sum_\ell\sigma^-(j\ell,t)\exp[-{\rm i}\mathbf k\cdot
\mathbf r(j\ell)].
\end{eqnarray*}
These conjugate variables may be calculated directly for each spin
configuration (indexed by the variable $t$). In particular, they do
not rely on any \emph{a priori} knowledge of the exchange integrals
$J$, nor the number and type of significant interactions for each
spin. We proceed by assembling the magnon variables into the column
vectors $\boldsymbol\tau(\mathbf k,t)$,
$\boldsymbol\tau^\ast(\mathbf k,t)$, from which the $t$-averaged
spin coordinate matrix is formed:
\begin{equation*}
\boldsymbol\Sigma(\mathbf k)=\langle\boldsymbol\tau^\ast(\mathbf
k)\cdot\boldsymbol\tau^{\rm T}(\mathbf k)\rangle.
\end{equation*}
The elements of $\boldsymbol\Sigma(\mathbf k)$ are given by
\begin{equation}
\begin{split}
\Sigma_{j,j^{\prime}}(\mathbf
k)=&\frac{\hbar}{2N}\sqrt{S_jS_{j^{\prime}}}\sum_{\ell\ell^{\prime}}\langle\sigma^-(j\ell)\sigma^+(j^{\prime}\ell^{\prime})\rangle\\&\quad\times\exp\{{\rm
i}\mathbf k\cdot[\mathbf r(j^{\prime}\ell^{\prime})-\mathbf
r(j\ell)]\}.
\end{split}
\end{equation}
The angular brackets used here represent an average taken either
over time (\emph{e.g.}, for spin dynamical (SD) simulations) or over
configurations (\emph{e.g.}, for ergodic ensembles). In practice,
some degree of error will necessarily be introduced in this process,
by considering finite time intervals and/or finitely large
configurational ensembles. This error diminishes with the square
root of the number of configurations or timesteps included in the
averaging process.

We proceed to investigate the properties of the matrix
$\boldsymbol\Sigma(\mathbf k)$ further, first forming its product with
the spin-dynamical matrix, $\boldsymbol\Delta(\mathbf
k)\cdot\boldsymbol\Sigma(\mathbf k)$. The diagonal entries of the
product matrix are given explicitly by:
\begin{equation}\label{ddots}
\begin{split}
[\boldsymbol\Delta(\mathbf k)\cdot\boldsymbol\Sigma(\mathbf
k)]_{j,j}=&\sum_{j^{\prime},\ell^{\prime}}S_jS_{j^{\prime}}J{j
j^{\prime}\choose
0\ell^{\prime}}[\langle\sigma^-(j0)\sigma^+(j0)\rangle\\&\quad-\langle\sigma^-(j^{\prime}\ell^{\prime})\sigma^+(j0)\rangle].
\end{split}
\end{equation}
The off-diagonal entries $[\boldsymbol\Delta(\mathbf
k)\cdot\boldsymbol\Sigma(\mathbf k)]_{i,j}$ ($i\ne j$) contain terms
whose exchange integrals $J(i,j^{\prime})$ and spin displacement
correlations $\langle\sigma^-(j^{\prime})\sigma^+(j)\rangle$ do not
share like terms; consequently, their time average vanishes. As such,
$\boldsymbol\Delta(\mathbf k)\cdot\boldsymbol\Sigma(\mathbf k)$ is a
diagonal matrix whose diagonal entries are given by Eq.\,\eqref{ddots}.

Within the high-temperature limit, we proceed to show that the
diagonal entries are equal to the thermal energy $k_{\rm B}T$. This
is seen by expanding the spin Hamiltonian dot product in terms of
the spin variables:
\begin{equation*}
\mathbf S_i\cdot\mathbf
S_j=S_iS_j\left[1+\tfrac{1}{2}(\sigma^+_i\sigma^-_j+\sigma^-_i\sigma^+_j-\sigma^+_i\sigma^-_i-\sigma^+_j\sigma^-_j)\right].
\end{equation*}
Here we have neglected fourth-order terms in $\sigma^\pm$ as---in
the limit of small fluctuations---one has
$\left|\sigma^\pm\right|\ll1$. Substituting this expansion back into
the Hamiltonian,
\begin{equation}\label{hamiltonianexpansion}
{\cal H}=-\sum_{i\ne j}J(i,j)S_iS_j+\sum_{i\ne
j}J(i,j)S_iS_j[\sigma_i^+\sigma_i^--\sigma_i^+\sigma_j^-].
\end{equation}
The first term on the right-hand side of this expression corresponds
to the spin interaction energy of the system in the absence of any
spin excitations, while the second term contains the correction that
arises from the population of spin-wave modes. The summation in this
spin-wave term runs over all non-trivial pairs $(i,j)$, and so can
be separated into contributions ${\cal H}_i$ from each spin $i$:
\begin{equation*}
{\cal H}=-\sum_{i\neq j}J(i,j)S_iS_j+\sum_i{\cal H}_i,
\end{equation*}
where
\begin{equation*}
{\cal
H}_i=\sum_jJ(i,j)S_iS_j[\langle\sigma_i^+\sigma_i^-\rangle-\langle\sigma_i^+\sigma_j^-\rangle].
\end{equation*}
Here, each Hamiltonian has been replaced by its time average, and so
the ${\cal H}_i$ give the time-averaged thermal energy of each spin
$i$; in the high-temperature limit this is simply $k_{\rm B}T$, by
the classical equipartition result. A more formal quantum-mechanical
treatment is given below; however, the classical result suffices
here. By comparison with Eq.\,\eqref{ddots}, one has
\begin{equation}
[\boldsymbol\Delta(\mathbf k)\cdot\boldsymbol\Sigma(\mathbf
k)]_{j,j}=k_{\rm B}T.
\end{equation}
This is a key result and gives that the matrix
$\boldsymbol\Delta(\mathbf k)\cdot\boldsymbol\Sigma(\mathbf k)$ is
diagonal with diagonal entries $k_{\rm B}T$.

One corollary of this result is that the eigenvalues of
$\boldsymbol\Sigma(\mathbf k)$ are the quantities $k_{\rm
B}T/\omega(\mathbf k,\nu)$ for each spin-wave mode $\nu$. This is an
important result because the matrix $\boldsymbol\Sigma(\mathbf k)$ is
constructed from the $\boldsymbol\tau(\mathbf k,t)$ and
$\boldsymbol\tau^\ast(\mathbf k,t)$, the values of which can be
determined directly from inspection of the spin displacements in spin
configurations. Consequently, the calculation and diagonalisation of
the spin coordinate matrix acts as a model-independent method of
extracting spin-wave frequencies from spin configurations.

\subsection{Implementation}

In order to assemble a set of spin-wave dispersion curves, one need
only calculate $\boldsymbol\Sigma(\mathbf k)$ for an appropriate range
of values of the wave-vector $\mathbf k$ (corresponding to, for
example, the particular reciprocal-space directions of interest). The
maximum ``resolution'' obtainable relies on the number of unit cells
represented by each configuration: a supercell containing $n_a,n_b,n_c$
unit cells along axes $\mathbf a,\mathbf b,\mathbf c$ permits the set
of wave-vectors:
\begin{equation*}
\mathbf k=\frac{i_a}{n_a}\mathbf a^\ast+\frac{i_b}{n_b}\mathbf
b^\ast+\frac{i_c}{n_c}\mathbf c^\ast\qquad i_a,i_b,i_c\in\mathbb{Z}.
\end{equation*}
Diagonalisation of $\boldsymbol\Sigma(\mathbf k)$ at each wave-vector
then gives the frequencies of the spin-wave modes at that point in
reciprocal space. But the method yields more in that the eigenvectors
of $\boldsymbol\Sigma(\mathbf k)$ describe the spin-displacement
patterns associated with each normal mode. In this way, the modes may
be individually labelled according to their particular displacement
pattern, and connected accordingly from $\mathbf k$-point to $\mathbf
k$-point to form the spin-wave dispersion curves. By calculating mode
frequencies at symmetry-equivalent wave-vectors, it is possible to
estimate the error associated with each point on the curve. Moreover,
the acoustic mode frequency at $\mathbf k=0$ gives a first-order
estimate of the configurational finite-size effects.

It would be possible to extend this analysis through consideration of
the implications of lattice symmetry on the form of
$\boldsymbol\Delta(\mathbf k)$ and $\boldsymbol\Sigma(\mathbf k)$. By
this we mean that one may in principle calculate first the normal mode
displacement vectors using standard symmetry arguments, and then use
these as a basis in which to express the observed spin displacements.
In this representation, the spin displacement matrix would be in
block-diagonal form, with each block corresponding to a particular
irreducible representation of the point group of the magnetic
structure. Individual blocks could then be separated and treated
individually, allowing unambiguous identification of the symmetry of
each of the normal modes. A similar approach has been applied elsewhere
to the analysis of phonon modes from atomistic
configurations.\cite{Tautz_1991,Goodwin_2005}

\subsection{Quantum-mechanical derivation of the ``spin coordinate matrix''}

We now recover the above results from a quantum-mechanical analysis,
using standard definitions and notation as described in \emph{e.g.}\
Ref.\,\onlinecite{Kittel_1987}. For convenience, this derivation is
limited to systems that contain only one spin type; however, the
formalisms developed are readily extended to include multiple spin
systems. It is easily shown that the spin Hamiltonian can be
expressed in terms of the standard magnon creation and annihilation
operators ${\rm a}^\ast,{\rm a}$:
\begin{equation*}
{\cal H}=-\sum_{i\ne j}\left[J(i,j)S^2+2J(i,j)S({\rm a}_i^\ast{\rm
a}_j-{\rm a}_i^\ast{\rm a}_i)\right].
\end{equation*}
This result holds only under quasi-saturation conditions, where the
spin deviations are small.\cite{quasisaturation} Noting the
similarity to Eq.\,\eqref{hamiltonianexpansion}, the first term on
the right-hand side of this expression represents the energy
contribution that arises from the spin alignment process itself
(\emph{i.e.}, in the absence of any spin-wave excitations); the
second represents the (positive) energy contribution due to thermal
excitations of the spins.

We now introduce the Holstein-Primakoff magnon
variables,\cite{Holstein_1940}
\begin{eqnarray*}
{\rm b}_{\mathbf k}&=&\frac{1}{\sqrt{N}}\sum_j{\rm a}_j\exp({\rm
i}\mathbf
k\cdot\mathbf r_j),\\
{\rm b}_{\mathbf k}^\ast&=&\frac{1}{\sqrt{N}}\sum_j{\rm
a}_j^\ast\exp(-{\rm i}\mathbf k\cdot\mathbf r_j),
\end{eqnarray*}
which are related to the creation and annihilation operators by reverse
Fourier transform:
\begin{eqnarray*}
{\rm a}_j&=&\frac{1}{\sqrt{N}}\sum_{\mathbf k}{\rm b}_{\mathbf
k}\exp(-{\rm i}\mathbf
k\cdot\mathbf r_j),\\
{\rm a}_j^\ast&=&\frac{1}{\sqrt{N}}\sum_{\mathbf k}{\rm b}_{\mathbf
k}^\ast\exp({\rm i}\mathbf k\cdot\mathbf r_j),
\end{eqnarray*}
and can be used as a basis in which to express the spin Hamiltonian.
Some manipulation gives
\begin{equation}
{\cal H}=-\sum_{i\ne j}J(i,j)S^2+\sum_{\mathbf k}\gamma_{\mathbf
k}\hbar{\rm b}_{\mathbf k}^\ast{\rm b}_{\mathbf
k},\label{quantumhamiltonian}
\end{equation}
where
\begin{equation*}
\gamma_{\mathbf k}=\frac{2S}{\hbar N}\sum_{i\ne
j}J(i,j)[1-\exp(-{\rm i}\mathbf k\cdot\mathbf r_{ji})]
\end{equation*}
and $\mathbf r_{ji}=\mathbf r_j-\mathbf r_i$. Again, it is the
second term on the right-hand side of
Eq.\,\eqref{quantumhamiltonian} that arises from the population of
spin-wave modes. This expression can in fact be described in terms
of the spin-dynamical and spin coordinate matrices as defined above.
The identities ${\rm a}_j=(S/2)^{1/2}\sigma^+$ and ${\rm
a}_j^\ast=(S/2)^{1/2}\sigma^-$ give
\begin{eqnarray*}
\hbar{\rm b}_{\mathbf k}^\ast{\rm b}_{\mathbf
k}&=&\frac{\hbar}{N}\sum_{i,j}{\rm a}_i^\ast{\rm a}_j\exp({\rm
i}\mathbf k\cdot\mathbf r_{ji})\nonumber\\
&=&\frac{\hbar S}{2N}\sum_{i,j}\sigma_i^-\sigma_j^+\exp({\rm
i}\mathbf k\cdot\mathbf r_{ji})\nonumber\\
&=&\boldsymbol\Sigma(\mathbf k).
\end{eqnarray*}
Moreover,
\begin{equation*}
\begin{split}
\gamma_{\mathbf k}&=-\frac{2S}{\hbar N}\sum_{j\ell\ne
j^{\prime}\ell^{\prime}}\left[J{jj^{\prime}\choose\ell\ell^{\prime}}\right.\\
&\left.\vphantom{J{jj^{\prime}\choose\ell\ell^{\prime}}}\qquad\times\biglb(\exp\{-{\rm
i}\mathbf k\cdot[\mathbf r(j^{\prime}\ell^{\prime})-\mathbf
r(j\ell)]\}-1\bigrb)\right]\\
&=\boldsymbol\Delta(\mathbf k),
\end{split}
\end{equation*}
by comparison with Eq.\,\eqref{deltak}. Consequently, the
Hamiltonian can in fact be written in the form
\begin{equation}\label{finalquantum}
{\cal H}=-\sum_{i\ne j}J(i,j)S^2+\sum_{\mathbf
k}\boldsymbol\Delta(\mathbf k)\cdot\boldsymbol\Sigma(\mathbf k).
\end{equation}

The spin-wave energy term in Eq.\,\eqref{finalquantum} represents the
contribution from thermal population of the individual modes (treating
modes at different wave-vectors separately). The final step in the
derivation arises from the quantum-mechanical result that the magnon
variables give the mode occupation number:\cite{Kittel_1987} the
eigenvalues of ${\rm b}_{\mathbf k}^\ast{\rm b}_{\mathbf k}$ are the
occupation numbers $n(\mathbf k)$. Consequently the eigenvalues of
$\boldsymbol\Sigma(\mathbf k)=\hbar{\rm b}_{\mathbf k}^\ast{\rm
b}_{\mathbf k}$ are the values $\hbar n(\mathbf k)\simeq k_{\rm
B}T/\omega(\mathbf k)$ (in the high temperature limit). This
quantum-mechanical analysis also suggests the diagonalised form of
$\boldsymbol\Sigma(\mathbf k)$ (which we denote as
$\boldsymbol\Omega(\mathbf k)$ and $\boldsymbol\Delta(\mathbf
k)\cdot\boldsymbol\Sigma(\mathbf k)$ when the high-temperature
approximations are not used:\label{quantumkbt}
\begin{equation}\label{boseeinsteinstatistics}
\boldsymbol\Omega(\mathbf k)=\hbar n(\mathbf
k)=\frac{\hbar}{\exp[\hbar\omega(\mathbf k)/k_{\rm B}T]-1},
\end{equation}
giving in turn
\begin{equation}
\boldsymbol\Delta(\mathbf k)\cdot\boldsymbol\Sigma(\mathbf
k)=\omega(\mathbf k)\boldsymbol\Omega(\mathbf
k)=\frac{\hbar\omega(\mathbf k)}{\exp[\hbar\omega(\mathbf k)/k_{\rm
B}T]-1}.\label{noht}
\end{equation}
We note that in the high-temperature limit, Eq.\,\eqref{noht}
reduces to the classical equipartition result cited earlier.

\section{Case study: Spin Monte-Carlo Simulation of a 3D $S=\frac{1}{2}$ Heisenberg
Ferromagnet}\label{casestudy}

In order to test our approach we prepared an ensemble of spin
configurations using spin Monte-Carlo simulations. Spin Monte-Carlo
is an ideal method for a test case such as this, as one has complete
control over the strength and nature of spin interactions, and hence
the spin dynamical information reflected in its output. Consequently
the form of the spin dispersion that should be recoverable from a
configurational ensemble is known precisely. For example, the
spin-wave dispersion for a 3D $S=\frac{1}{2}$ Heisenberg ferromagnet
is easily shown to be
\begin{equation}\label{3dhf}
\hbar\omega(\mathbf k)=4SJ[3-\cos(ak_x)-\cos(ak_y)-\cos(ak_z)].
\end{equation}
Spin Monte-Carlo simulations in which pairs of spins interact via
classical ferromagnetic Heisenberg potentials should then yield
configurations that reflect this same dispersion behaviour.

\begin{figure} \centering
\includegraphics[width=8.0cm]{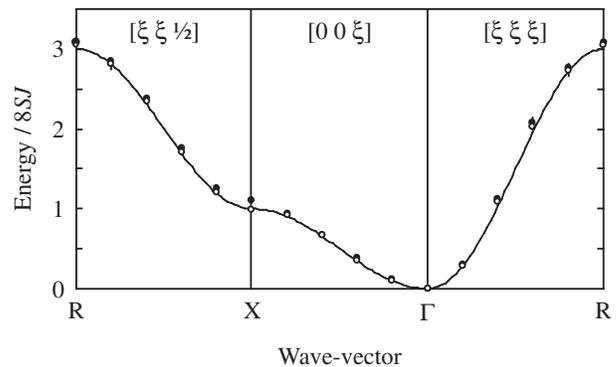}
\caption{\label{fig1}Spin-wave dispersion curves in a 3D Heisenberg
ferromagnet calculated using Eq.\,\eqref{3dhf} (solid line), and
extracted from our spin Monte-Carlo simulations using the method
developed in the text (open circles---10\,mK; filled
circles---1\,K).}
\end{figure}

With this in mind, we generated an ensemble of \emph{ca} 500
equilibrium Monte-Carlo configurations, each representing a
$10\times10\times10$ supercell of an idealised primitive cubic
ferromagnet. The simulation employed a coupling constant
$J=0.4$\,meV and was carried out at two temperatures below its
paramagnetic phase transition temperature $T_{\rm c}\simeq3.5$\,K:
10\,mK and 1\,K. We analysed our configurations according to the
method described above along the symmetry directions
$[\xi\xi\frac{1}{2}]$, $[00\xi]$ and $[\xi\xi\xi]$ at a resolution
of 0.1 reciprocal lattice units. In our analysis, it was necessary
to account for the classical population statistics inherent to
Monte-Carlo simulations, rather than the Bose-Einstein populations
implied by Eq.\,\eqref{boseeinsteinstatistics}. The first-order
finite-size effects given by the zone-centre spin-wave frequencies
were subtracted from the raw data; these were very small for the
$T=1$\,K data, and essentially non-existent for the lower
temperature set. The error associated with each point on the
dispersion curves was estimated by comparison of the energy values
obtained at symmetry-equivalent wave-vectors. Our results are
compared with the theoretical dispersion curves in Fig.\,\ref{fig1}.

The excellent agreement between observed (Monte-Carlo generated spin
configuration, analysed via the spin coordinate matrix) and
calculated [Eq.~\eqref{3dhf}] spin-wave dispersion is strong
evidence for the applicability of the theoretical approach described
in this paper. In this test case, we simply recover the very
information that we used to drive the Monte-Carlo
simulations---namely the value of $J$ and the nature of the spin
interactions. But the key result is that these quantities were
extracted \emph{solely} from a set of individual spin orientations.
The analysis is blind to the method with which these were generated,
and indeed the particular interaction parameters employed. We have
shown that these parameters may be recovered quantitatively from
atomistic configurations.

\section{Discussion and Conclusions}

Our general approach could be automated very easily in the form of a
computer program, and our limited experience of implementing the
analysis in this way has shown that the calculation of spin-wave
dispersion curves of relatively simple systems can be carried out
using a conventional desktop computer over very manageable ($<1$\,h)
timescales. The generation of sufficiently many configurations to
yield results with acceptable error margins will in general demand
significantly more extensive computational resources. We have found
that for simple systems, one requires in the vicinity of 500
independent configurations to give acceptable results.

When analysing SD simulations, one has access to more than spin
orientations alone, and the use of this additional information can help
improve the quality of the spin-wave dispersion curves obtained. Given
that the simulations calculate changes in orientation as a function of
real time, it is possible to calculate the individual spin momenta
across a given configuration. Our method of analysis may be extended by
calculating---in addition to the spin displacement matrix
$\boldsymbol\Sigma(\mathbf k)$---the related spin momentum matrix
$\dot{\boldsymbol\Sigma}(\mathbf k)$, constructed from the momentum
variables $\dot{\sigma}^\pm(j\ell,t)$. Then it is easily shown that the
ratio of the eigenvalues $\dot{e}(\mathbf k,\nu)$ of
$\dot{\boldsymbol\Sigma}(\mathbf k)$ to the eigenvalues $e(\mathbf
k,\nu)$ of $\boldsymbol\Sigma(\mathbf k)$ is related to the spin-wave
mode frequencies:
\begin{equation}\label{momentumexpression}
\omega^2(\mathbf k,\nu)=\frac{\dot{e}(\mathbf k,\nu)}{e(\mathbf
k,\nu)}.
\end{equation}
While computationally more intensive, the calculation of spin-wave
frequencies in this manner yields results of a higher quality,
particularly in the form of the long-wavelength acoustic
modes.~\cite{Dove_1993} The origin of this improvement is particular
to the nature of SD simulations, where the timescale sampled during
the simulation might not be sufficient to ensure equipartition of
energy between all normal modes. The formalism of
Eq.\,\eqref{momentumexpression} ensures that one evaluates the
extent of spin displacement in the context of the population of a
particular spin-wave mode, rather than the expected population as
given by the overall extent of spin-wave excitation. This extension
is of course irrelevant in the analysis of ergodic simulations,
where there is no real concept of time.

In conclusion, we have shown how a relatively straightforward
extension of standard spin-wave theory can be used to obtain a
quantitative link between the energies of the various spin-wave
modes in a material and the individual spin orientations one might
observe in ``atomistic'' models of its behaviour. The method we
describe has the particular advantage that it allows the extraction
of spin-dynamical quantities without any prior assumption of the
form of the various spin interactions. This model-independence would
be of significant advantage when using the technique to study
materials for which little is known \emph{a priori} about the nature
of the magnetic interactions, or where one wishes to avoid the
presumption of a particular interaction model.

\section*{Acknowledgements}

We acknowledge financial support from the EPSRC (U.K.), and from
Trinity College, Cambridge to A.L.G. We thank Dr A.\ T.\ Boothroyd
for helpful discussions.


\begin{references}
\bibitem{quasisaturation}
An analogous approximation was used in the quasi-classical treatment
given above, when assuming $|\sigma^\pm|\ll1$.
\end{references}
\end{document}